\documentclass[%
 reprint,
 amsmath,amssymb,
 aps,
]{revtex4-1}

\usepackage{graphicx}
\usepackage{dcolumn}
\usepackage{bm}

\usepackage{accents}
\newcommand{\dbtilde}[1]{\accentset{\approx}{#1}}

\usepackage{amsbsy}
\usepackage{amssymb}

\newcommand{\I}{\mathrm{i}} 
\newcommand{\D}{\mathrm{d}} 
\newcommand{\E}{\mathrm{e}} 

\begin{document}

\title{Charged massive vector boson propagator \\
in a constant magnetic field in arbitrary $\xi$-gauge \\
obtained using the modified Fock-Schwinger method}

\author{Iablokov S. N.}
\email{physics@iablokov.ru}
\affiliation{%
 P.G. Demidov Yaroslavl State Univeristy, Yaroslavl, Russia \\
 A.A. Kharkevich Institute for Information Transmission Problems, Moscow, Russia
}%

\author{Kuznetsov A. V.}%
\affiliation{%
 P.G. Demidov Yaroslavl State Univeristy, Yaroslavl, Russia
}%

\date{\today}

\begin{abstract}
We applied a recently published modified Fock-Schwinger (MFS) method to find the exact solution of the propagator equation for a charged vector boson in the presence of a constant magnetic field directly in the momentum space as a sum over Landau levels in arbitrary $\xi$-gauge.  
In contrast to the standard approaches for finding propagators, MFS method demonstrated several improvements in terms of computational complexity reduction and revealed simple internal structures in intermediate and final expressions, thus allowing to obtain new useful representations of the propagator.
\end{abstract}

\pacs{Valid PACS appear here}
\maketitle


\section{\label{sec:level1}Introduction}

Analysis of elementary particle loop processes in extreme conditions, such as strong magnetic fields, requires a knowledge of particle propagators where the field effects are taken into account exactly. 

There exist at least two naturally arising scales of strong magnetic fields. The first one corresponds to the so-called critical, or Schwinger value $B_e = m_e^2/e \simeq 4.4 \times 10^{13}$ G, which is the strength of the quantizing field for an electron (hereafter, we use the Planck units: $\hbar = 1, c = 1$). The fields of the order of $B_e$ are connected with the concept of magnetars, i.e. neutron stars which evolution is driven largely by magnetic fields~\cite{Duncan:1992}. 
Other examples when such strong (and even stronger) magnetic fields could possibly manifest themselves include the experiments at modern colliders, e.g. with non-central collisions of heavy ions~\cite{Skokov:2009}, and high-intensity electromagnetic waves generated by a system of lasers~\cite{Korzhimanov_2011,DiPiazza:2012,Tajima:2012}.

The second scale is defined by the mass of the gauge boson $m_W$: $B_W = m_W^2/e \simeq 1.1 \times 10^{24}$~G~\cite{Grasso:2001}. In this case, there arises a question of applicability of the Standard Model in these conditions, namely, the stability of electroweak vacuum at $B\rightarrow B_{W}$. As it was shown in Ref.~\cite{Skalozub:2014}, the radiation corrections act to prevent the instability of the electroweak vacuum in such strong fields. 

A knowledge of the vector-boson propagator at the scale of $B_W$ expanded over the Landau levels can be helpful for 
investigations of processes in the early Universe. An example of possible influence of the quantizing effect of the strong magnetic field on the $W$ propagator was considered in Ref.~\cite{Kuznetsov:2017}. A model was used of dynamical generation of the primordial magnetic field in the early Universe by ferromagnetic domain walls, see~\cite{Campanelly:2006} and the references cited therein. 
Due to this effect, the decay width of $\nu_e\to e^{-}W^{+}$ has a ``sawtooth'' profile, thus leading to the significant decrease of neutrino mean free path at some neutrino energies. If the lepton-antilepton asymmetry (induced by the $CP$ violation in the lepton sector) has arisen before the electroweak phase transition, leading to overabundance of neutrinos over antineutrinos in the Universe, then the considered mechanism would provide an overabundance of $W^+$ over $W^-$ inside domain walls. The subsequent decay of the $W$ boson by dominant quark channels could have influenced the baryon asymmetry in the early Universe. 


Quantum field propagators can be either constructed as time-ordered correlation functions of the field operators or found from the propagator equation provided by the path integral formalism. The first approach requires to solve the corresponding field equation in order to construct a quantum field operator. The solutions should be normalized and, for fields with spin, orthogonalized with respect to some spin operator. As an intermediate computational step, the spin parts should be multiplied and summed over. However, these spin-related manipulations are absent in the second approach. The solution of the propagator equation already contains all the spin parts summed over, and the $\delta$-function in the right-hand side ensures the correct normalization. Therefore, the latter approach  seems to require less computational effort to obtain a propagator.

Additional difficulties arise when some external field, e.g. electromagnetic, is present. In this case, one can no longer rely on the translational invariance when applying Fourier transform. In some scenarios (e.g. constant external magnetic field), this could be remedied through the use of the Fock-Schwinger (FS) proper-time method \cite{Schwinger:1951, Itzykson_1980}. However, this approach leads to the expression for the propagator as an integral over the proper-time parameter. In order to get the expression in the momentum space as a sum over Landau levels, which is convenient for the calculation of scattering amplitudes, one can apply integration techniques described in \cite{KM_Book_2013}.

In this paper, we applied a recently published modified Fock-Schwinger (MFS) approach \cite{MFS_2018} to the solution of the massive vector boson propagator equation in the presence of a constant magnetic field in arbitrary $\xi$-gauge. It allowed to obtain the propagator expression directly in the momentum space as a sum over the Landau levels. The paper is structured as follows. First, we briefly mention known results and approaches for finding quantum field propagators in external electromagnetic fields, and provide a known expression for the case of W-boson, however, obtained using the East-Coast metric convention. Next, we discuss the main steps of the MFS method and use it to get the expression for the massive vector boson propagator in the same metric convention. Finally, we apply the above-mentioned integration techniques \cite{KM_Book_2013} to the proper-time representation expression of the propagator to show that, in East-Coast metric convention, the transformed expression coincides with the result obtained using the MFS approach. In the end, we also derive the propagator equation in the West-Coast metric convention (commonly used in modern particle physics) and provide the corresponding solution.


\section{\label{sec:level1}Known approaches and results}

A history of calculations of charged particle propagators in a magnetic field is rather long. 
The exact expression for the electron propagator in a constant uniform magnetic field was first obtained by J. Schwinger~\cite{Schwinger:1951} in the Fock-Schwinger~\cite{Fock:1937} proper time formalism. There exist a number of papers where another forms of the propagator are derived. 
For example, the case of a superstrong magnetic field was analyzed in Ref.~\cite{Loskutov:1976} where the contribution from the ground Landau level to the electron propagator was obtained. 
In Ref.~\cite{Chodos_1990}, the propagator was transformed from the Schwinger form~\cite{Schwinger:1951} to a series over an integer number $n$, 
where the poles in the expansion terms corresponded to Landau levels. An exact proof for the propagator~\cite{Chodos_1990} 
to be an expansion over the Landau levels, was presented in Ref.~\cite{Kuznetsov:2011_Okr}, 
where the electron propagator in a constant uniform magnetic field was obtained in the same form using the exact solutions of the Dirac equation.
A misprint in the formula for the propagator~\cite{Chodos_1990} was later corrected in Refs.~\cite{Gusynin:1999, Chyi_2000}, but without any comments.
In Ref.~\cite{Chyi_2000}, the expansion of the electron propagator as a power series of the intensity of a magnetic field was presented. 

The formula for the propagator of a charged scalar particle expanded over Landau levels was obtained for the first time in Ref.~\cite{Ayala:2005}. 

The propagator equation (and the corresponding solution) for $W$-boson in a constant magnetic field for an arbitrary $\xi$-gauge was previously obtained in \cite{Erdas_1990} (see also \cite{Erdas_2000}) using the East-Coast metric convention $g_{\mu\nu} = (-,+,+,+)$, and was given by:
\begin{equation}
\label{eq_g_east}
H^\mu_{\,\,\, \nu} G^\nu_{\,\, \rho}(X,X') = \delta^\mu_{\,\,\, \rho}  \delta^{(4)}(X-X') \, ,
\end{equation}
where
\begin{equation}
\label{h_east}
H^\mu_{\,\,\, \nu} = \left( \Pi \Pi + m^2 \right) \delta^\mu_{\,\,\, \nu} - 2ieQF^\mu_{\,\,\, \nu} + \left( \frac{1}{\xi} - 1  \right) \Pi^\mu \Pi_\nu  \, .
\end{equation}
The following standard notations are also assumed:
\begin{eqnarray}
\nonumber
X^\mu &=& (t, x, y, z) \, , \quad \, \, \, \, \, X_\mu = g_{\mu\nu} X^\nu = (-t, x, y, z) \, , \quad \quad
\\
\nonumber
\partial_\mu &=& (\partial_t, \nabla) \, , \quad \quad \quad \,  \partial^\mu = (-\partial_t, \nabla) \, ,
\\
\nonumber
\label{convention_east}
D_\mu &=& \partial_\mu + \I e Q A_\mu \, , \, \, \, \, \, \Pi_\mu = - \I D_\mu \, , \quad \,\, \Pi\Pi = \Pi^\nu \Pi_\nu \, .
\end{eqnarray}

Here, $Q$ is a dimensionless charge of the $W^{(-)}$ particle, which is equal to $-1$, and $e>0$ is the elementary charge. 
Electromagnetic field configuration
\begin{equation}
\label{eq_7a}
\begin{gathered}
F^{\mu\nu} = \partial^\mu A^\nu - \partial^\nu A^\mu
\end{gathered}
\end{equation}
for a constant magnetic field along the $z$-axis is given by $F^{12} = - F^{21} = B$, with the rest of components being equal to $0$.
The solution of Eq.~(\ref{eq_g_east}) was obtained using the Fock--Schwinger method and, in proper-time representation, reads: 
\begin{equation}
\label{w_prop_erdas}
\begin{gathered}
G^\mu_{\,\, \nu}(X,X') = \phi(X, X') \int \frac{\D^4 p}{(2\pi)^4} \E^{\I (p(X-X'))} G^\mu_{\,\, \nu}(p) \, ,
\end{gathered}
\end{equation}
with the translationally non-invariant phase factor
\begin{equation}
\label{w_prop_erdas_phi}
\begin{gathered}
\phi(X,X') = \exp{\bigg(- \frac{\I e Q}{2} X'^{\rho} F_{\rho\sigma} X^\sigma \bigg) },
\end{gathered}
\end{equation}
and the Fourier transform of the translationally invariant part
%
\onecolumngrid
\vspace{\columnsep}
\begin{widetext}
\begin{eqnarray}
\label{w_prop_erdas_fourier}
G^\mu_{\,\, \nu}(p) & =& \I \int^\infty_0 \frac{\D s}{\cos(\beta s)} \E^{-\I s \left( p^2_\parallel + p^2_\perp \frac{\tan (\beta s)}{\beta s} \right)} 
\bigg\{ 
\E^{-\I s \left( m^2 - \I \varepsilon \right)} \left[ \delta^\mu_{\parallel \nu} + \delta^\mu_{\perp \nu} \cos (2\beta s)  - Q \, \varphi^\mu_{\,\,\, \nu} \sin (2\beta s) \right] \, + 
\\[2mm]
&+&\frac{1}{m^2} \left( \E^{-\I s \left( m^2 - \I \varepsilon \right)} - \E^{-\I s \left( \xi m^2 - \I \varepsilon \right)} \right) 
 \bigg[ \bigg( p^\mu - Q \, \left(\varphi p\right)^\mu \, \tan \left(\beta s\right) \bigg) \bigg( p_\nu - Q \,\left(p \varphi \right)_\nu \, \tan \left(\beta s\right) \bigg) -
\nonumber
\\[2mm]
&-& \frac{\I \beta}{2} \bigg( Q \, \varphi^\mu_{\,\,\, \nu} + \delta^\mu_{\perp \nu} \tan (\beta s) \bigg) \bigg] 
\bigg\} \, . \nonumber
\end{eqnarray}
\end{widetext}
\vspace{\columnsep}
\twocolumngrid
Here, the subscript $\perp$ stands for the components orthogonal to the direction of the magnetic field, namely, belonging to the plane $(x, y)$ for the field directed along the $z$-axis, while the subscript $\parallel$ stands for $t$ and $z$ components in this case.
We also introduced the dimensionless magnetic field tensor $\varphi^\mu_{\,\,\, \nu} = F^\mu_{\,\,\, \nu} /B$ and an auxiliary notation $\beta = e B$.
The tensor indices of four-vectors and tensors standing inside the parentheses are contracted consecutively, e.g. 
$\left(\varphi p\right)^\mu = \varphi^\mu_{\,\,\, \lambda} \, p^\lambda$. 

Alternatively, in Ref.~\cite{Nikishov}, the W-boson propagator was constructed in the Feynman gauge ($\xi = 1$) as a time-ordered product of the field operators that were expanded as a series over the solutions of the corresponding wave-equation. 
Finally, in Ref.~\cite{KuznetsovOkruginShitova:2015} (see also \cite{KM_Book_2013}), the proper-time parameter $s$ in (\ref{w_prop_erdas_fourier}) was integrated out, giving the Fourier transform that coincided (for $\xi = 1$) with the result of Ref.~\cite{Nikishov}.


Knowledge of different representations of the charged particle propagators 
in an external magnetic field is important because it allows to consider the conditions of their applicability. 
There exist several precedents when misunderstanding of such conditions
led to incorrect studies. For instance, a calculation of the neutrino self-energy operator in a magnetic field 
was performed in Refs.~\cite{Elizalde:2002,Elizalde:2004} by analyzing the one-loop diagram
$\nu \to e^-\, W^+ \to \nu$. The authors restricted themselves
by the contribution to the electron propagator from the ground Landau level.
As it was shown in Ref.~\cite{Kuznetsov:2006}, in that case the contribution 
from the ground Landau level did not dominate due to the large electron virtuality, 
and contributions from other levels were of the same order. Ignoring
such a fact led the authors~\cite{Elizalde:2002,Elizalde:2004} to incorrect results. 
Another example of this kind was an attempt to reanalyze
the probability of the neutrino decay $\nu \to e^- W^+$ in an external magnetic field
in the limit of ultra-high neutrino energies, calculated via the imaginary part
of the one-loop amplitude of the transition $\nu \to e^- \, W^+ \to \nu$.
Initially, the result was obtained in Ref.~\cite{Erdas:2003}. Later, the calculation was repeated in
~\cite{Bhattacharya:2009} where authors insisted on another
result. The third independent calculation \cite{Kuznetsov:2010_PLB}
confirmed the result of Ref.~\cite{Erdas:2003}. The most likely cause of the
error in Ref.~\cite{Bhattacharya:2009} was that the authors used only linear terms in the expansion of the $W$-boson propagator over
the electromagnetic tensor $F^{\mu\nu}$ whereas the quadratic terms were essential as well.


The modified Fock-Schwinger approach, developed in \cite{MFS_2018} and discussed below, provides additional useful
representations of the W-boson propagator.

\section{\label{sec:level1}Outline of the modified Fock-Schwinger approach}
Here we present a brief overview of the modified Fock-Schwinger (MFS) approach. We are to solve the following propagator equation:
\begin{equation}
\label{eq_hg_delta}
H(\partial_X, X) \, G(X, X') = \delta^{(4)} (X-X') \, .
\end{equation}
As in the original Fock-Schwinger (FS) method (see, e.g., \cite{Itzykson_1980}) one should, first, switch to the following integral representation:
\begin{equation}
\label{q}
G(X, X') = \I \int_{0}^{\infty} \D s \, U(X,X';s) \, .
\end{equation}
Considering $U(X,X';s)$ as some sort of an evolution operator satisfying a Schr{\"o}dinger-type equation
\begin{equation}
\label{eq_3}
\I \, \partial_s \, U(X, X'; s) = H(\partial_X, X) \, U(X, X'; s) 
\end{equation}
with the appropriate boundary conditions
\begin{eqnarray}
\label{u_boundary}
U(X,X';\infty) &=& 0 \, ,
\\
\nonumber
U(X,X';0) &=& \delta^{(4)} (X-X') \,,
\end{eqnarray}
one obtains the following result:
\begin{eqnarray}
\label{u_exp_delta}
U(X, X'; s) &=& \E^{ -\I s \left[ \, H(\partial_X, X) - \I \varepsilon \right] } \, \delta^{(4)} (X-X') \, ,
\\
\label{g_inv_h_delta}
G(X, X') &=& H^{-1}(\partial_X, X) \, \delta^{(4)} (X-X') \, ,
\end{eqnarray}
where
\begin{equation}
\label{h_inverse}
H^{-1}(\partial_X, X) = \I \int_{0}^{\infty} \D s \, \E^{ -\I s \left[ \, H(\partial_X, X) - \I \varepsilon \right] } \,
\end{equation}
is the inverse of $H$.

The $\I \varepsilon$ prescription was added in order to satisfy the boundary conditions (\ref{u_boundary}). From now on, we will skip writing it explicitly, always assuming its presence.  

Expressions, such as (\ref{u_exp_delta}), considered in the framework of the distribution theory, make perfect sense due to the infinite differentiability of the $\delta$-function. 
In the original FS method, one reduces the task of finding $U$ to a solution of a special differential equation. The MFS approach, however, consists in the direct   evaluation of the exponential operator action on the $\delta$-function. In order to do so, an appropriate representation of the $\delta$-function should be chosen:
\begin{equation}
\label{eq_delta_representation}
\delta^{(4)} (X-X') = \sum \int \psi_\lambda(X) \psi_\lambda(X') \, ,
\end{equation}
where $\psi_\lambda(X)$ is an eigenvector of the $H$ operator:
\begin{equation}
\label{eigen_h}
H(\partial_X, X) \psi_\lambda(X) = H(\lambda) \psi_\lambda(X) \, .
\end{equation}
Therefore, Eq.~(\ref{g_inv_h_delta}) simplifies to:
\begin{equation}
\label{eq_exp_tau}
G(X, X') = \I \int_{0}^{\infty} \D s \, \sum \int \E^{ -\I s \, H(\lambda) }  \psi_\lambda(X) \psi_\lambda(X') \, .
\end{equation}
Next, the exponential part is integrated out:
\begin{equation}
\label{eq_7b}
G(X, X') =  \sum \int \frac{\psi_\lambda(X) \psi_\lambda(X')}{ H(\lambda)} \, .
\end{equation}
In many cases, it is not $H$ itself that satisfies (\ref{eigen_h}), but rather a part of it:
\begin{equation}
\label{eq_7c}
\begin{gathered}
H = H_0 + H_1 \, , \\
H_0(\partial_X, X) \psi_\lambda(X) = H_0(\lambda) \psi_\lambda(X) \, .
\end{gathered}
\end{equation}
\\
If $H_0$ commutes with $H_1$, that indeed is the case for the problem discussed below, the exponential operator decomposes into two parts and the solution takes the following form:
\begin{equation}
\label{eq_7d}
G(X, X') = \I \int_{0}^{\infty} \D s \, \sum \int \E^{ -\I s \, H_1 } \E^{ -\I s \, H_0(\lambda) }  \psi_\lambda(X) \psi_\lambda(X') \, .
\end{equation}
Further simplifications highly depend on the exact form of $H_0$ and $H_1$. However, in some cases it is possible to transform corresponding expressions
such that the dependence on $s$ is accounted through the exponential factor as in (\ref{eq_exp_tau}), which allows for straightforward evaluation of
the integral over $s$.

\section{\label{sec:level1} $W$-boson propagator in a constant magnetic field \\ \quad \quad \quad (East-Coast metric)}

Let's apply MFS method to the equation (\ref{eq_g_east}) using the East-Coast metric convention $(-,+,+,+)$ throughout this section. First, we notice that the left-hand side operator $H$ consists of three parts ($H = H_0 + H_F + H_\xi$):
\begin{eqnarray}
\label{h0_east}
\left( H_0 \right)^\mu_{\,\,\, \nu} \, &=& \left( \Pi \Pi + m^2 \right) \delta^\mu_{\,\,\, \nu} \, ,
\\
\label{eq_H_F}
\left(H_F \right)^\mu_{\,\,\, \nu} \, &=& - 2ieQF^\mu_{\,\,\, \nu} \, ,
\\
\label{eq_H_xi}
\left(H_\xi \right)^\mu_{\,\,\, \nu} \, &=& \left( \frac{1}{\xi} - 1  \right) \Pi^\mu \Pi_\nu \, .
\end{eqnarray}
There exist several useful commutation relations for the case of a constant electromagnetic field $F$:
\begin{eqnarray}
\label{eq_comm_east}
\left[\Pi^\mu, \Pi^\nu \right] &=& -\I eQF^{\mu\nu} \, ,
\\
\left[\Pi^\mu, \Pi \Pi \right] &=& -2\I eQF^\mu_{\,\,\, \nu} \Pi^\nu \, ,
\nonumber \\
\left[ \Pi^\mu \Pi_\nu, \Pi \Pi \right] &=& 
-2\I eQ  \left( F^\mu_{\,\,\, \rho} \Pi^\rho \Pi_\nu - \Pi^\mu \Pi_\rho F^\rho_{\,\,\, \nu} \right)
\, .
\nonumber
\end{eqnarray}
Only one of three commutators between parts of $H$ is vanishing ($\left[H_0, H_F\right] = 0$). 
Two others are not equal to zero separately, however, their sum is. This leads to $[H_0 + H_F, H_\xi] = 0$, thus, allowing for a step-by-step separation of $\E^{-\I s \left( H_0 + H_F + H_\xi \right)}$:
\begin{equation}
\label{eq_7e}
\begin{gathered}
\E^{-\I s \left( H_0 + H_F + H_\xi \right)} = \E^{-\I s H_\xi} \, \E^{-\I s \left( H_0 + H_F \right)} = \\
\E^{-\I s H_\xi} \, \E^{-\I s H_F} \, \E^{-\I s H_0} .
\\
\end{gathered}
\end{equation}
\\
Let's briefly discuss the anatomy of so constructed propagator:
\begin{equation}
\label{anatomy}
G(X, X') = \I \int_{0}^{\infty} \D s \, \E^{-\I s H_\xi} \, \E^{-\I s H_F} \, \E^{-\I s H_0} \, \delta^{(4)} (X-X') \, .
\end{equation}
The $H_0$ part is the basic building block, which represents the propagation of a scalar particle. The $H_F$ part adds some additional structure due to the spin properties of a vector boson, similar to the case of electron's propagator \cite{MFS_2018}. However, for $\xi \neq 1$ there is yet another layer of complexity due to the choice of $\xi$-gauge. 
In order to proceed with further calculations, we choose the Landau gauge for the electromagnetic potential $A^\mu$:
\begin{equation}
\label{eq_a_mu}
A^\mu = (0, 0, Bx, 0) \, .
\end{equation}
Making a standard change of variables
\begin{equation}
\label{def_eta}
\eta = \sqrt{\beta} \left( x + Q \frac{p_y}{\beta} \right) \, , \quad \eta ' = \sqrt{\beta} \left( x' + Q \frac{p_y}{\beta} \right) \, ,
\end{equation}
we consider the following $\delta$-function representation (the same as in \cite{MFS_2018}):
\begin{equation}
\label{delta_repr}
\delta^{(4)} (X-X') = \sqrt{\beta} \sum_{n=0}^\infty \int \frac{\D ^3 p_{\shortparallel, y}}{(2\pi)^3} \,
\E^{{\I \left( p(X-X') \right)}_{\shortparallel, y}} V_n V_n ' \, .
\end{equation}
Here, $V_n = V_n(\eta) $ [$V_n ' = V_n(\eta ')$] is a shorthand notation for the $n$-th level quantum harmonic oscillator (QHO) eigenfunction:
\begin{equation}
\label{eq_7f}
V_n(\eta) = \frac{\E^{-\eta^2/2} H_n(\eta)}{\sqrt{2^n \, n! \, \sqrt{\pi}}} \, ,
\end{equation}
where $H_n$ is a Hermite polynomial.
Accounting for the action of the operator ($\ref{h0_east}$) on the $\E^{\I (px)}$-type expressions, we obtain:
\begin{equation}
\label{eq_7g}
\left( H_0 \right)^\mu_{\,\,\, \nu} \, = \left( p^2_\parallel + m^2 - \beta(\partial^2_\eta -\eta^2) \right) \delta^\mu_{\,\,\, \nu} \, .
\end{equation}
This form of $H_0$, therefore, justifies the $\delta$-function representation (\ref{delta_repr}) due to the following equation for QHO eigenfunctions:
\begin{equation}
\label{eq_7h}
(\partial^2_\eta -\eta^2)V_n = -(2n+1)V_n \, .
\end{equation}
This being said, we evaluate the action of the first exponential operator in Eq.~(\ref{anatomy}):
\onecolumngrid
\vspace{\columnsep}
\begin{widetext}
\begin{eqnarray}
\label{eq_result_exp_h_0}
\left( \E^{-\I s H_0} \right)^\mu_{\,\,\, \nu}
 \, \delta^{(4)} (X-X') = \sqrt{\beta} \sum_{n=0}^\infty \int \frac{\D ^3 p_{\shortparallel, y}}{(2\pi)^3} \,
\E^{-\I s \left[ p^2_\parallel + m^2 +(2n+1)\beta \right]} \E^{{\I \left( p(X-X') \right)}_{\shortparallel, y}} V_n V_n ' \; 
\delta^\mu_{\,\,\, \nu} \, .
\end{eqnarray}
Expanding the exponential series for the operator $H_F$, see Eq.(\ref{eq_H_F}), one obtains the following expression:
\begin{equation}
\label{e_h_f}
\left[\E^{-\I s H_F}\right]^\mu_{\,\,\, \nu} = 
\left[\E^{-2 Q \beta s \varphi}\right]^\mu_{\,\,\, \nu} = 
\delta^\mu_{\parallel  \nu}  + \cos (2\beta s) \delta^\mu_{\perp  \nu} - Q \sin (2\beta s) \varphi^\mu_{\,\,\, \nu} =
\delta^\mu_{\parallel  \nu} + \frac{\E^{\I 2\beta s}}{2}\left( \delta^\mu_{\perp  \nu} + \I Q\varphi^\mu_{\,\,\, \nu} \right) + \frac{\E^{-\I 2\beta s}}{2}\left( \delta^\mu_{\perp  \nu} - \I Q\varphi^\mu_{\,\,\, \nu} \right) .
\end{equation}
Next, we shift the summation index over the Landau levels, such that the expression in the exponent stays the same for all terms in Eq.~(\ref{eq_result_exp_h_0}). 
This gives the expression for the consecutive action of two exponential operators on the $\delta$-function:  
\begin{eqnarray}
\label{eq_result_exp_h_f}
&& \left( \E^{-\I s H_F} \E^{-\I s H_0} \right)^\mu_{\,\,\, \nu}
 \, \delta^{(4)} (X-X') = \sqrt{\beta} \sum_{n=-1}^\infty \int \frac{\D ^3 p_{\shortparallel, y}}{(2\pi)^3}  \,
\E^{{\I \left( p(X-X') \right)}_{\shortparallel, y}} \, \E^{-\I s \left[ p^2_\parallel + m^2 +(2n+1)\beta \right]} \, d^\mu_{\,\,\, \nu} \, ,
\\
\label{d1_east}
&& d^\mu_{\,\,\, \nu} = \delta^\mu_{\parallel  \nu} V_n V_n ' + 
\frac{1}{2}\left( \delta^\mu_{\perp  \nu} + \I Q\varphi^\mu_{\,\,\, \nu} \right) V_{n+1} V_{n+1} ' + 
\frac{1}{2}\left( \delta^\mu_{\perp  \nu} - \I Q\varphi^\mu_{\,\,\, \nu} \right) V_{n-1} V_{n-1} '  \, .
\end{eqnarray}
\end{widetext}
\vspace{\columnsep}
\twocolumngrid
%

Finally, the exponential operator for the gauge-dependent part $H_\xi$ simplifies to:
\begin{equation}
\label{eq_7i}
\E^{-\I s \left( \frac{1}{\xi} - 1 \right) \Pi^\mu \Pi_\nu } = \delta^\mu_{\,\,\, \nu} + \Pi^\mu \frac{ \E^{-\I s \left( \frac{1}{\xi} - 1 \right) \Pi \Pi  } - 1 }{\Pi\Pi} \Pi_\nu \, .
\end{equation}
In the MFS approach, the gauge-dependent part decomposes into the sum of two terms. The first term represents the choice of Feynman gauge ($\xi=1$) and leads to the expression given by Eqs.~(\ref{eq_result_exp_h_f}), (\ref{d1_east}). The second one accounts for additional complexity in the case $\xi \neq 1$.
\\
\\
In order to evaluate the expression
\begin{equation}
\label{xi_neq_1_action}
\bigg[ \Pi^\mu \frac{ \E^{-\I s \left( \frac{1}{\xi} - 1 \right) \Pi \Pi  } - 1 }{\Pi\Pi} \Pi_\nu \bigg] \E^{-\I s H_F} \E^{-\I s H_0} \, \delta^{(4)} (X-X')
\end{equation}
let's introduce some standard notations:
\begin{eqnarray}
a^+ &=& \frac{1}{\sqrt{2}} (\eta - \partial_\eta) \, ,  \quad \quad  a^- = \frac{1}{\sqrt{2}} (\eta + \partial_\eta)  \, ,
\\
\partial_\eta &=& \frac{1}{\sqrt{2}} (a^- - a^+) \, ,  \quad \, \, \, \,  \eta = \frac{1}{\sqrt{2}} (a^- + a^+) \, .
\end{eqnarray}
The following auxiliary vector
\begin{equation}
\label{eq_7j}
v_\rho = \left( 0, \frac{i}{\sqrt{2}}, \frac{Q}{\sqrt{2}}, 0 \right)
\end{equation}
is also useful for further computations due to its properties:
\begin{equation}
\label{eq_7k}
\begin{gathered}
v_\rho \left( \delta^\rho_{\perp \nu} + \I Q \varphi^\rho_{\,\,\, \nu} \right) = 0  \, , \quad \quad
v_\rho^* \left( \delta^\rho_{\perp \nu} + \I Q \varphi^\rho_{\,\,\, \nu} \right) = 2 v_\nu^* \, ,
\\
v_\rho^* \left( \delta^\rho_{\perp \nu} - \I Q \varphi^\rho_{\,\,\, \nu} \right) = 0  \, , \quad \quad 
v_\rho \left( \delta^\rho_{\perp \nu} - \I Q \varphi^\rho_{\,\,\, \nu} \right) = 2 v_\nu \, .
\end{gathered}
\end{equation}
Therefore, the action of 
\begin{equation}
\label{eq_7l}
\Pi_\rho = p_{\parallel \, \rho} + \sqrt{\beta} v_\rho a^+ + \sqrt{\beta} v^*_\rho a^-
\end{equation}
on $d^\rho_{\,\,\, \nu}$ is given by:
\begin{eqnarray}
\label{Pi_d}
\Pi_\rho d^\rho_{\,\,\, \nu} &=& p_{\parallel \, \nu} V_n V_n ' +
\\
\nonumber
&+& \sqrt{n\beta} \, v_\nu \, V_n V_{n-1} ' + \sqrt{(n+1)\beta} \, v^*_\nu \, V_n V_{n+1} ' \,.
\end{eqnarray}
We notice that all QHO eigenfunctions that depend on $\eta$ have the same index $n$, thus, justifying the following substitution in expression (\ref{xi_neq_1_action}):
\begin{equation}
\label{eq_7m}
\Pi \Pi \to p^2_\parallel + (2n+1) \beta \, .
\end{equation}
Therefore, the middle part of [...] operator in Eq.~(\ref{xi_neq_1_action}) can be moved away safely as a $c$-number, even prior to the evaluation of $\Pi^\mu$ action on expression (\ref{Pi_d}). This results in the following intermediate formula:
\onecolumngrid
\vspace{\columnsep}
\begin{widetext}
\begin{eqnarray}
 \left( \E^{-\I s H} \right)^\mu_{\,\,\, \nu}
 \, \delta^{(4)} (X-X') &=&
 \left( \E^{-\I s H_\xi} \, \E^{-\I s H_F} \, \E^{-\I s H_0} \right)^\mu_{\,\,\, \nu} \, \delta^{(4)} (X-X') =  \sqrt{\beta} \sum_{n=-1}^\infty \int \frac{\D ^3 p_{\shortparallel, y}}{(2\pi)^3} \, \times
\\  
&\times& \left( \delta^\mu_{\,\,\, \rho} + \frac{\E^{-\I s \left( \frac{1}{\xi} - 1 \right) \left[ p^2_\parallel +(2n+1)\beta \right]} - 1}{p^2_\parallel +(2n+1)\beta} \Pi^\mu \Pi_\rho \right) 
\E^{{\I \left( p(X-X') \right)}_{\shortparallel, y}} \, \E^{-\I s \left[ p^2_\parallel + m^2 +(2n+1)\beta \right]} \, d^\rho_{\,\,\, \nu} \, .
\nonumber
\end{eqnarray}
Now that the dependence on $s$ is brought to the exponents, one could easily integrate it out:
\begin{eqnarray}
\label{g_py_repr}
G^\mu_{\,\,\, \nu}(X, X') &=& \I \int_{0}^{\infty} \D s \, \E^{ -\I s H } \, \delta^{(4)} (X-X') = 
\\
&=& \sqrt{\beta} \sum_{n=-1}^\infty \int \frac{\D ^3 p_{\shortparallel, y}}{(2\pi)^3}  \frac{\E^{{\I \left( p(X-X') \right)}_{\shortparallel, y}}}{p^2_\parallel + m^2 + (2n+1)\beta}
\left( d^\mu_{\,\,\, \nu} + \frac{\xi - 1}{p^2_\parallel + \xi m^2 + (2n+1)\beta} f^\mu_{\,\,\, \nu} \right) \, ,
\nonumber
\\
f^\mu_{\,\,\, \nu} &\equiv& \Pi^\mu \Pi_\rho d^\rho_{\,\,\, \nu} \, = p^{\,\mu}_{\parallel} p^{\,}_{\parallel \, \nu} V_{n} V_{n}' \, +
\\  
&+& p^{\,\mu}_{\parallel} \left( \sqrt{\beta n} \, v_\nu V_{n} V_{n-1}' + \sqrt{\beta (n+1)} \, v^*_\nu V_{n} V_{n+1}' \right) \, + 
\left( \sqrt{\beta n} \, v^{*\mu} V_{n-1} V_{n}' + \sqrt{\beta (n+1)} \, v^{\mu} V_{n+1} V_{n}' \right) \,p^{\,}_{\parallel  \nu}  + 
\nonumber \\
&+& \beta \sqrt{n(n+1)} \left( v^{\mu} v_{\nu} V_{n+1} V_{n-1}' + v^{*\mu} v^{*}_{\nu} V_{n-1} V_{n+1}' \right) \, +
\beta (n+1) v^{\mu} v^{*}_{\nu} V_{n+1} V_{n+1}' + \beta n \, v^{*\mu} v_{\nu} V_{n-1} V_{n-1}' \, .
\nonumber
\end{eqnarray}
\end{widetext}
\vspace{\columnsep}
\twocolumngrid
Among many explicit representations of the propagator, expression (\ref{g_py_repr}) is not the most useful one.
It is asymmetric with respect to $x,y$ coordinates. In order to symmetrize it, let's evaluate the integral over $\D p_y$. 
We notice that the integrand depends on $p_y$ not just through the exponential factor but also through $\eta$ and $\eta '$ in QHO functions, see Eq.~(\ref{def_eta}).
Therefore, the following integrals for different $n$ and $n'$ should be calculated:
\begin{equation}
\label{eq_7.1}
\begin{gathered}
I_{n,n'} = \int \D p_y e^{\I p_y(y-y')} V_n(\eta) V_{n'} (\eta') \, .
\end{gathered}
\end{equation}
First, we make a change of the integration variable:
\begin{equation}
\label{eq_7.2}
\begin{gathered}
u = Q \frac{p_y}{\sqrt{\beta}} + \frac{\sqrt{\beta}}{2} \left[ (x+x') - \I Q (y-y') \right] \, .
\end{gathered}
\end{equation}
This leads to:
\begin{equation}
\label{eq_7.3}
\begin{gathered}
I_{n,n'} = \frac{\E^{\I \Phi(X,X')}}{\sqrt{2^{n+n'} n! n'! \, \pi}} \sqrt{\beta} \, \E^{-\frac{\beta}{4} (X-X')^2_\perp} \tilde{I}_{n,n'} \, ,
\end{gathered}
\end{equation}
where
\begin{eqnarray}
\label{eq_7.4}
\Phi(X,X') &=& - \frac{Q\beta}{2} (x+x')(y-y') \, ,
\\
\tilde{I}_{n,n'} &=& \int_{-\infty}^\infty \D u \, \E^{-u^2} H_n (u+a) H_{n'} (u+b) \, ,
\end{eqnarray}
with the following substitutions:
\begin{eqnarray}
a &=& \frac{\sqrt{\beta}}{2} \left[ (x-x') + \I Q (y-y') \right] \, ,
\\
b &=& - \frac{\sqrt{\beta}}{2} \left[ (x-x') - \I Q (y-y') \right] \, .
\nonumber 
\end{eqnarray}
The phase~(\ref{eq_7.4}) can be shown to be in agreement with the one in Eq.~(\ref{w_prop_erdas_phi}),
see e.g. Ref.~\cite{KM_Book_2013}.
Secondly, according to Ref.~\cite{Gradshteyn_Ryzhik}, the $\tilde{I}_{n,n'}$ integral evaluates to:
\begin{equation}
\label{eq_7.5}
\begin{gathered}
\tilde{I}_{n,n'} = 2^{n'} \sqrt{\pi} \, n! \, b^{n'-n} L^{(n'-n)}_{n}(-2ab) = 
\\
= 2^{n'} \sqrt{\pi} \, n! \, b^{n'-n} L^{(n'-n)}_{n}\left(\frac{\beta}{2}Z^2_\perp \right)  
 \quad \quad \left[n \leq n' \right] \, ,
\end{gathered}
\end{equation}
where the functions $L^{(m)}_n$ are the Laguerre polynomials and $Z^\mu = X^\mu - X'^{\mu}$ .
The symmetrized representation of the propagator then reads:
\onecolumngrid
\vspace{\columnsep}
\begin{widetext}
\begin{eqnarray}
G^\mu_{\,\,\, \nu}(X, X') &=& \frac{\beta}{2 \pi} \E^{\I \Phi} \sum_{n=-1}^\infty \int \frac{\D^2 p_{\shortparallel}}{(2\pi)^2} \frac{\E^{{\I \left( pZ \right)}_{\parallel}} \, \E^{-\beta Z^2_\perp / 4}}{p^2_\parallel + m^2 + (2n+1)\beta}
\left( \tilde{d}^\mu_{\,\,\, \nu} + \frac{\xi - 1}{p^2_\parallel + \xi m^2 + (2n+1)\beta} \tilde{f}^\mu_{\,\,\, \nu} \right) \, ,
\\
\label{d2_east}
\tilde{d}^\mu_{\,\,\, \nu}  &=& \delta^\mu_{\parallel \nu} L_n + \frac{1}{2} \delta^\mu_{\perp \nu} \bigg ( L_{n+1} + L_{n-1} \bigg) + \frac{\I Q}{2} \varphi^\mu_{\,\,\, \nu} \bigg( L_{n+1} - L_{n-1} \bigg) \, ,
\\
\label{f2_east}
\tilde{f}^\mu_{\,\,\, \nu}  &=& \bigg[ p^\mu_\parallel p_{\parallel \nu} - \frac{\beta Q}{2} \bigg( p^\mu_\parallel 
(Z \varphi)_\nu + (\varphi Z)^\mu p_{\parallel \nu} \bigg) + \bigg( (2n+1)\beta -\frac{\beta^2}{4}Z^2_\perp \bigg) \delta^\mu_{\perp \nu} + \frac{\I Q \beta}{2} \, \varphi^\mu_{\,\,\, \nu}  \bigg] L_n + \, 
\\
\nonumber
&+& \frac{\I \beta}{2} \bigg[ \bigg( p^\mu_\parallel Z_{\perp \nu} + Z^\mu_\perp p_{\parallel \nu} \bigg) + \I \delta^\mu_{\perp \nu} - \frac{Q \beta}{2} Z^2_\perp \varphi^\mu_{\,\,\, \nu} \bigg] \bigg( L^{(1)}_n + L^{(1)}_{n-1} \bigg) 
- \beta^2 (\varphi Z)^\mu (Z \varphi)_\nu L^{(2)}_{n-1} \, .
\end{eqnarray}
In (\ref{d2_east}) and (\ref{f2_east}), all the Laguerre polynomials $L_n^{(m)}$ have $\beta Z_\perp^2 / 2$ as their arguments: $L_n^{(m)} = L_n^{(m)} (\beta Z_\perp^2 / 2$).
This form of the propagator is only a partial momentum space representation (with respect to $t,z$ coordinates). However, the established symmetry allows
to calculate the remaining Fourier transforms (with respect to $x,y$ coordinates). In order to do this, one should switch to polar coordinates in the $x-y$ plane and make use of the integral identities \cite{Gradshteyn_Ryzhik}:
\begin{equation}
\label{gradshteyn_rhyzhik_1}
\begin{gathered}
\int^{2\pi}_0 \D\varphi \, \E^{-\I \zeta cos(\varphi) \pm \I m \varphi} = (-\I)^m \, 2\pi J_m(\zeta) \, ,
\end{gathered}
\end{equation}
\begin{equation}
\label{gradshteyn_rhyzhik_2}
\begin{gathered}
\int^{\infty}_0 \D \zeta \, \zeta^{\lambda / 2} \E^{-p\zeta}  J_\lambda(b\sqrt{\zeta}) L^{(\lambda)}_n(c\zeta) 
= \left( \frac{b}{2} \right)^\lambda \frac{(p-c)^n}{p^{\lambda+n+1}} \E^{-\frac{b^2}{4p}} L^{(\lambda)}_n \left( \frac{b^2c}{4pc-4p^2} \right) \, ,
\end{gathered}
\end{equation}
where $J_m$ are the Bessel functions of order $m$.
After some simple but rather cumbersome rearrangements and substitutions, one finally arrives to the following result:

\begin{eqnarray}
\label{g_east}
G^\mu_{\,\,\, \nu}(X, X') &=& \E^{\I \Phi} \int \frac{\D^4 p}{(2\pi)^4} \E^{\I \left( p (X-X') \right)} G^\mu_{\,\,\, \nu}(p) \, ,
\\
\label{g_p_east}
G^\mu_{\,\,\, \nu}(p) &=& \sum_{n=-1}^\infty \frac{ 2 (-1)^n \, \E^{-p^2_\perp / \beta}}{p^2_\parallel + m^2 + (2n+1)\beta}
\left( \dbtilde{d}^\mu_{\,\,\, \nu} + \frac{\xi - 1}{p^2_\parallel + \xi m^2 + (2n+1)\beta} \dbtilde{f}^\mu_{\,\,\, \nu} \right) \, ,
\\
\label{d3_east} 
\dbtilde{d}^\mu_{\,\,\, \nu} &=& \delta^\mu_{\parallel \nu} L_n - \frac{1}{2} \delta^\mu_{\perp \nu} \left( L_{n+1} + L_{n-1} \right) - \frac{\I Q}{2} \varphi^\mu_{\,\,\, \nu} \left( L_{n+1} - L_{n-1} \right) \, ,
\\
\label{f3_east}
\dbtilde{f}^\mu_{\,\,\, \nu} &=& \left( p^\mu p_\nu - \frac{\I Q \beta}{2} \varphi^\mu_{\,\,\, \nu} \right) L_n - 
(\varphi p)^\mu (p \varphi)_\nu \left( L_n + 4 L^{(2)}_{n-1} \right) \, +
\\
&& + \, \I Q \left( p^\mu (p \varphi)_\nu + (\varphi p)^\mu p_\nu + \frac{\I Q \beta}{2} \delta^\mu_{\perp \nu} \right) \left( L^{(1)}_n + L^{(1)}_{n-1} \right) \, .
\nonumber
\end{eqnarray}
In (\ref{d3_east}) and (\ref{f3_east}), all the Laguerre polynomials $L_n^{(m)}$ have $2 p_\perp^2 / \beta$ as their arguments: $L_n^{(m)} = L_n^{(m)} (2 p_\perp^2 / \beta$).


\section{\label{sec:level1} Comparison of two ways of obtaining the $W$-boson propagator}

Let's transform Eq.~(\ref{w_prop_erdas}) according to the recipe from \cite{Chodos_1990} (see also \cite{KM_Book_2013} and \cite{Chyi_2000}) and then compare the result with Eqs.~(\ref{g_east}) - (\ref{f3_east}). The Fourier transform of the translationally invariant part of the $W$ boson propagator~(\ref{w_prop_erdas}) in the $\xi$-gauge is presented in Eq.~(\ref{w_prop_erdas_fourier}). Let us rewrite it in a more convenient form:
\begin{eqnarray}
\label{w_prop_erdas_transformed_1}
G^\mu_{\,\,\, \nu}(p) 
&=& \frac{\I}{\beta}\int\limits_0^{\infty}\D v\, \E^{-\I\rho v}\, 
 \biggl[ \delta^\mu_{\parallel \nu} \, F_1 (v) + \delta^\mu_{\perp \nu} \, F_2 (v) - Q \, \varphi^{\mu}_{\,\,\, \nu} \, F_3 (v) \biggr] + \frac{\I}{\beta m^2} \int \limits_0^{\infty} \D v \left( \E^{-\I \rho v} - \E^{-\I \rho_\xi v} \right) \times
\nonumber \\[2mm]
&\times& 
\biggl[ 
\biggl( p^\mu p_\nu - \I \, \frac{Q \beta}{2} \, \varphi^{\mu}_{\,\,\, \nu} \biggr)\, F_1 (v) \, 
- \biggl( Q \, p^\mu (p \varphi)_\nu  + Q (\varphi p)^\mu p_\nu + \I \, \frac{\beta}{2} \delta^\mu_{\perp \nu}  \biggr) 
\, F_4 (v) + (\varphi p)^\mu (p \varphi)_\nu \, F_5 (v) \,
\biggr] \, ,
\end{eqnarray}
\end{widetext}
\vspace{\columnsep}
\twocolumngrid

where the functions are introduced:
\begin{eqnarray}
\label{f_functions}
F_1 (v) &=& \frac{1}{\cos v} \, \exp (- \I \, \alpha \, \tan v) \,, \\
F_2 (v) &=& \frac{\cos(2\,v)}{\cos v} \, \exp (- \I \, \alpha \, \tan v) \,, \\
F_3 (v) &=& \frac{\sin(2\,v)}{\cos v} \, \exp (- \I \, \alpha \, \tan v) \,, \\
F_4 (v) &=& \frac{\tan v}{\cos v} \, \exp (- \I \, \alpha \, \tan v) = \I \frac{\partial}{\partial \alpha} \, F_1 (v) \,, \\
F_5 (v) &=& \frac{\tan^2 v}{\cos v} \, \exp (- \I \, \alpha \, \tan v) = - \frac{\partial^2}{\partial \alpha^2} \, F_1 (v) \,, 
\end{eqnarray}
and $v = \beta s$, $\rho = ( m^2+ p_{\|}^2 )/\beta$, $\rho_\xi = ( \xi \, m^2+ p_{\|}^2 )/\beta$, 
$\alpha = p_{\perp}^2 /\beta$. 

Since all the functions $F_j (v) \, (j = 1...5)$ satisfy the relation $F_j (v+\pi\,n) = (-1)^n\,F_j (v)$, let us
divide the integration domain $(0, \infty)$ into intervals 
$(0, \pi), (\pi, 2 \pi), \dots , (n \pi, [n+1] \pi), \dots$. Making in each segment the change 
of variable, $v \to v + n \pi$, we can write:
\begin{eqnarray}
\label{f_exp_integrals}
I_j &=& \int\limits_0^{\infty} \D v \, \exp (- \I \rho v) \, F_j (v) = 
\nonumber \\
&=& \sum\limits_{n=0}^\infty (-1)^n\, \exp (- \I \rho n \pi) A_j  = 
\\
&=&  \frac{1}{1 + \exp (- \I \rho \pi)} \; A_j \,,
\nonumber 
\end{eqnarray}
where
\begin{eqnarray}
A_j = \int\limits_0^{\pi} \D v \, \exp (- \I \rho v) \, F_j (v) \,.
\label{eq:S(q)_A_j}
\end{eqnarray}

The details of calculations of the functions $A_j$ can be found in Refs.~\cite{KM_Book_2013, KuznetsovOkruginShitova:2015}. One finally obtains:
\begin{eqnarray}
\label{eq:intpg1}
A_1 &=& -2\,\I \left( 1 + \E^{- \I \rho \pi} \right) 
\sum\limits_{n = 0}^{\infty} \, \frac{\ell_{n-1}(\alpha)}{\rho+2\,n - 1}\,,
\label{eq:intpg2}
\end{eqnarray}
\begin{eqnarray}
A_2 &=& -\I \left( 1 + \E^{- \I \rho \pi} \right)  \sum\limits_{n = 0}^{\infty} 
\frac{\ell_n(\alpha) + \ell_{n-2}(\alpha)}{\rho+2\,n - 1}\,,\\[5mm]
\label{eq:intpg3}
A_3 &=& - \left( 1 + \E^{- \I \rho \pi} \right) 
\sum\limits_{n = 0}^{\infty} \, \frac{\ell_n(\alpha) - \ell_{n-2}(\alpha)}{\rho+2\,n - 1}\,,\\[5mm]
\label{eq:intA7}
A_4 &=& 2\left( 1 + \E^{- \I \rho \pi} \right) 
\sum\limits_{n = 0}^{\infty} \, \frac{\ell'_{n-1}(\alpha)}{\rho+2\,n - 1}\,,\\[5mm]
\label{eq:intA8}
A_5 &=& 2\,\I \left( 1 + \E^{- \I \rho \pi} \right) 
\sum\limits_{n = 0}^{\infty} \, \frac{\ell''_{n-1}(\alpha)}{\rho+2\,n - 1}\,,
\end{eqnarray}
where the auxiliary functions were introduced to write the resulting expressions in a more compact form:
\begin{equation}
\ell_n(\alpha) = (-1)^n \E^{-\alpha} L_n (2 \alpha)\,.
\label{eq:ell_n_alpha}
\end{equation}

After substituting the integrals~(\ref{eq:intpg1})--(\ref{eq:intA8}) into the expression for the propagator and performing 
simple manipulations with Laguerre polynomials, we observe that (\ref{w_prop_erdas_transformed_1}) transforms exactly to (\ref{g_p_east}) - (\ref{f3_east}).


\section{\label{sec:level1} W-boson propagator in a constant magnetic field \\ \quad \quad \quad (West-Coast metric)}
Expressions (\ref{g_east}) -- (\ref{f3_east}) were obtained in the East-Coast metric convention $g_{\mu\nu} = (-, +, +, +)$, however, in modern particle
physics literature the West-Coast metric convention $g_{\mu\nu} = (+, -, -, -)$ prevails. In this section, the following standard notations are assumed:
\begin{eqnarray}
\label{convention_west}
\nonumber
X^\mu &=& (t, x, y, z) \, , \quad \, \, \, \, \, X_\mu = g_{\mu\nu} X^\nu = (t, -x, -y, -z) \, , \quad \quad
\\
\nonumber
\partial_\mu &=& (\partial_t, \nabla) \, , \quad \quad \quad \,  \partial^\mu = (\partial_t, - \nabla) \, ,
\\
\nonumber
D_\mu &=& \partial_\mu + \I eQA_\mu \, , \, \, \, \, \, \Pi_\mu = \I D_\mu \, , \quad \,\, \Pi\Pi = \Pi^\nu \Pi_\nu \, .
\end{eqnarray}
In order to get the expression for the propagator, we, first,
consider relevant terms in the Standard Model Lagrangian:
\begin{eqnarray}
\label{sml_3terms}
\mathcal{L} &=& \mathcal{L}_{W} + \mathcal{L}_{WWA} + \mathcal{L}_{WWAA} + \mathcal{L}_{gauge} \, ,
\\
\label{sml_w}
\mathcal{L}_{W} &=& - \frac{1}{2} W^{\dagger}_{\mu\nu} W^{\mu\nu} + m^2 \, W^{\dagger}_{\mu} W^{\mu} \, ,
\nonumber \\
\label{sml_wwa}
\mathcal{L}_{WWA} &=& \I e Q \bigg[ F^{\mu\nu} W^{\dagger}_{\mu} W_{\nu} - W^{\dagger}_{\mu\nu} A^{\mu} W^{\nu}  + W^{\mu\nu} A_{\mu} W^{\dagger}_{\nu} \bigg] \, ,
\nonumber \\
\label{sml_wwaa}
\mathcal{L}_{WWAA} &=& e^2 \bigg[ A^{\mu} W^{\dagger}_{\mu} A^{\nu} W^{\dagger}_{\nu}  - A^{\mu}A_{\mu} W^{\dagger}_{\nu} W^{\nu} \bigg] \, ,
\nonumber \\
\label{sml_w}
\nonumber
\mathcal{L}_{gauge} &=& - \frac{1}{\xi} \bigg( D_\mu W^\mu \bigg)^{\dagger} \bigg( D_\nu W^\nu \bigg) \, .
\end{eqnarray}
The last term accounts for gauge fixing, in the same way as in Ref.~\cite{Erdas_1990}. Relative signs in the Standard Model Lagrangian were
chosen according to the convention adopted by the majority of modern textbooks (see, Ref.~\cite{Signs, Langacker}). This implies the choice of $\eta = 1$ in 
\begin{equation}
W^{i}_{\mu\nu} = \partial_\mu W^{i}_{\nu} - \partial_\nu W^{i}_{\mu} - \eta g \varepsilon_{ijk} W^{j}_{\mu}W^{k}_{\nu} \, .
\end{equation}
However, in Ref.~\cite{Erdas_1990} this choice was made according to Ref.~\cite{Cheng_Li} with $\eta = -1$, thus leading to different relative signs in the Lagrangian and the corresponding wave equation.

Next, from the given Lagrangian we derive a field equation for W-boson in external electromagnetic field. 
Adding $\delta$'s to the right-hand side results in the corresponding propagator equation:
\begin{eqnarray}
\label{eq_w_west}
&& H^\mu_{\,\,\, \nu} G^\nu_{\,\, \rho}(X,X') = \delta^\mu_{\,\,\, \nu}  \delta^4(X-X') \, ,
\\
&& H^\mu_{\,\,\, \nu} = \left( \Pi \Pi - m^2 \right) \delta^\mu_{\,\,\, \nu} - 2ieQF^\mu_{\,\,\, \nu} + \left( \frac{1}{\xi} - 1  \right) \Pi^\mu \Pi_\nu  \, .
\nonumber
\end{eqnarray}

Finally, using the same procedure as in section IV, one obtains the expression for the propagator in the West-Coast metric convention:
\onecolumngrid
\vspace{\columnsep}
\begin{widetext}
\begin{eqnarray}
\label{g_west}
G^\mu_{\,\,\, \nu}(X, X') &=& \E^{-\I \Phi} \int \frac{\D^4 p}{(2\pi)^4} \E^{- \I \left( p (X-X') \right)} G^\mu_{\,\,\, \nu}(p) \, ,
\\
\label{g_p_west}
G^\mu_{\,\,\, \nu}(p) &=& \sum_{n=-1}^\infty \frac{ 2 (-1)^n \, \E^{-p^2_\perp / \beta}}{p^2_\parallel - m^2 - (2n+1)\beta}
\left( \dbtilde{d}^\mu_{\,\,\, \nu} + \frac{\xi - 1}{p^2_\parallel - \xi m^2 - (2n+1)\beta} \dbtilde{f}^\mu_{\,\,\, \nu} \right) \, ,
\\
\label{d3_west} 
\dbtilde{d}^\mu_{\,\,\, \nu} &=& \delta^\mu_{\parallel \nu} L_n - \frac{1}{2} \delta^\mu_{\perp \nu} \left( L_{n+1} + L_{n-1} \right) + \frac{\I Q}{2} \varphi^\mu_{\,\,\, \nu} \left( L_{n+1} - L_{n-1} \right) \, ,
\end{eqnarray}

\begin{eqnarray}
\label{f3_west}
\dbtilde{f}^\mu_{\,\,\, \nu} &=& \left( p^\mu p_\nu - \frac{\I Q \beta}{2} \varphi^\mu_{\,\,\, \nu} \right) L_n - 
(\varphi p)^\mu (p \varphi)_\nu \left( L_n + 4 L^{(2)}_{n-1} \right) \, -
\\
&& - \, \I Q \left( p^\mu (p \varphi)_\nu + (\varphi p)^\mu p_\nu + \frac{\I Q \beta}{2} \delta^\mu_{\perp \nu} \right) 
\left( L^{(1)}_n + L^{(1)}_{n-1} \right) \, .
\nonumber
\end{eqnarray}
\end{widetext}
\vspace{\columnsep}
\twocolumngrid


\section{\label{sec:level1} Discussion}
We applied the modified Fock-Schwinger (MFS) method to obtain the exact analytical solution for the propagator of the W-boson in external constant magnetic field in arbitrary $\xi$-gauge. Commutation relations (\ref{eq_comm_east}) allowed us to study the solution's internal structure by considering the consecutive action of exponential operators inside the inverse operator (\ref{anatomy}) for the propagator equation (\ref{eq_g_east}). Each of the exponential operators accounted for the contribution to the final expression due to the (i) propagation of charged particle, (ii) its spin interaction with magnetic field and (iii) the choice of $\xi$-gauge. A particular form of $\delta$-function decomposition reduced the action of these exponentials either to the eigenvalue substitution or to the use of the QHO ladder operators. Several straightforward manipulations with special functions allowed us to make a transition from partial Fourier transform (with respect to $t$, $y$ and $z$ coordinates) to the full Fourier transform, however, for the translationally invariant part of the propagator.

In contrast to the nowadays standard approaches for finding propagators, the MFS approach demonstrates several improvements in terms of computational complexity reduction and reveals simple internal structures in intermediate and final expressions. First of all, it does not require the full knowledge of exact analytical solution for the corresponding wave equation, but only a form of the solution for its scalar part ($H_0$). The remaining complexity (e.g., correct normalization) is hidden inside the appropriate decomposition of the $\delta$-function, while the commutation relations (\ref{eq_comm_east}) justify the use of the same decomposition for the full problem (with spin and gauge-dependent parts).  Secondly, if compared with the original Fock--Schwinger approach followed by application of integration techniques discussed in section IV, MFS method requires significantly less effort to evaluate the integral over the proper-time parameter, therefore, leading directly to a (partial) Fourier transform. Finally, the introduction of QHO eigenfunctions and ladder operators on early computational stages allows us to use the convenient relations from the QHO problem.

Provided that the form of the solution for the $H_0$ part is known and the corresponding commutation relations for the remaining parts of $H$ are satisfied, MFS approach has a potential for its efficient use in higher dimensions due to the structure of the inverse operator, where each exponential operator accounts for a specific layer of complexity. This makes it promising for further development of quantum field theory methods in external fields.
 

\,

\acknowledgments

We are grateful to A.~Ayala, A.\,Ya.~Parkhomenko and D.\,A.~Rumyantsev for useful remarks. 

The reported study was funded by RFBR, project number \mbox{19-32-90137}. 

 

\bibliography{MFS}

\end{document}